\documentclass{elsarticle}

\usepackage{atlasphysics} 
\usepackage{subfigure}
\usepackage{mathrsfs}
\usepackage[hyperindex,breaklinks]{hyperref}

\def\ptTrack{\ensuremath{p_{\mathrm{T}}^{\rm\; track}}}
\def\ntrk{\ensuremath{N_{\mathrm{trk}}}}

\journal{Physics Letters B}

\begin{document}
\def\@draftversion{}
\newcommand{\draftversion}[3]%
{
  \def\@draftversion{#1}
}
\newcommand{\draftheading}%
{
  %
  %
  \newcount\timehh\newcount\timemm
  \timehh=\time 
  \divide\timehh by 60 \timemm=\time
  \count255=\timehh\multiply\count255 by -60 \advance\timemm by \count255
  \markboth%
  {\hspace*{\fill}{\protect\small\bf \fbox{DRAFT}}
    \hspace*{\fill}
    \protect\makebox[0pt][r]{\protect\small\sl\today\ -- 
      \ifnum\timehh<10 0\fi\number\timehh\,:\,\ifnum\timemm<10 0\fi\number\timemm}
  }
  {\protect\makebox[0pt][l]{\protect\small\sl\today\ -- 
      \ifnum\timehh<10 0\fi\number\timehh\,:\,\ifnum\timemm<10 0\fi\number\timemm}
    \hspace*{\fill}
    {\protect\small\bf \fbox{DRAFT}}
    \hspace*{\fill}}
  \pagestyle{myheadings}
}

\begin{frontmatter}


\title{Search for Strong Gravity Signatures in Same-sign Dimuon Final States using the ATLAS detector\\ at the LHC}
\author{The ATLAS Collaboration}
\date{\today}
\draftversion{1.2}


\begin{abstract}
A search for microscopic black holes has been performed in a same-sign dimuon final
state using 1.3 $\ifb$ of  proton-proton collision data collected with the ATLAS 
detector at a centre of mass energy of 7 $\tev$ at the CERN Large Hadron Collider.  
The data are found to be consistent with the expectation from the Standard Model and
the results are used to derive exclusion contours in the context of a low scale gravity model.
\end{abstract}


\end{frontmatter}



  \renewcommand{\thefootnote}{\fnsymbol{footnote}}

\section{Introduction}
\label{sec:intro}
Models introducing extra dimensions can provide a solution to the hierarchy problem, the fact that the Planck scale $M_\mathrm{Pl} \sim
10^{16}$~TeV is much larger than the electroweak scale.  In some models of extra dimensions, the gravitational field can propagate
into  $(n+4)$-dimensions, where $n$ is the number of extra dimensions, while the Standard Model particles are restricted to four-dimensional space-time. Therefore, the gravitational field as measured in four dimensions  is reduced in strength from the 
fundamental gravitational field. As a result, the Planck scale in $(n+4)$-dimensions $\MD$ would be much smaller than 
the Planck scale in four dimensions $M_\mathrm{Pl}$, and possibly comparable to the electroweak scale.  
An example of such a model of extra dimensions is the ADD model, which is a model of large flat extra dimensions~\cite{ArkaniHamed:1998rs,Antoniadis:1998ig,ArkaniHamed:1998nn}.

If extra dimensions exist and \MD{} is in the  \TeV{} range, microscopic black holes with masses at the \TeV{} scale
could be produced at the Large Hadron Collider~\cite{Argyres:1998qn,Banks:1999gd,Dimopoulos:2001hw,Giddings:2001bu,Kanti:2004nr}.
Black holes are expected to be produced when the classical impact parameter of  two colliding partons is smaller than the
higher-dimensional horizon radius corresponding to a black hole with mass equal to the invariant mass of the colliding parton 
system. This paper considers higher-dimensional Schwarzschild solutions, as well as Kerr solutions for 
black holes with initial angular momentum equal to the relative angular momentum between the two colliding partons; parton
spin is ignored~\cite{Dai:2007ki}.

The production of black holes at the LHC would occur with a continuous mass distribution ranging
from approximately the reduced Planck scale $\MD$ to the proton-proton centre of mass energy of 7~TeV. The classical approximations 
used for black hole production and the semi-classical approximations for decay are expected to be valid only for masses well above the higher-dimensional
Planck scale. A lower threshold $\MTH$  is thus applied to the black hole mass to reduce the contributions
from regions where the models are invalid. The production cross section is set to zero if the parton-parton centre of 
mass energy is below  $\MTH$.

Once produced, a black hole starts to evaporate in a manner described  by Hawking radiation~\cite{Hawking:1974sw}
which determines the energy and multiplicity distributions of the emitted particles.  
The relative multiplicities of the emitted particles are
determined by the number of degrees of freedom of each particle type and the decay modes of emitted unstable particles.
Black hole events should therefore have a high multiplicity of high-$\pt$ particles which is the characteristic
feature exploited in this analysis. Models with rotating and non-rotating black holes are considered in this paper. The multiplicity of
high-$\pt$ particles is lower for rotating black holes~\cite{bhrot}.
No graviton initial-state radiation or emission from the black hole is considered. As a result of the emission 
of Hawking radiation, the mass of the produced black hole decreases. When the mass of the black hole approaches $\MD$,
quantum gravity effects become important.   In the final stage of the black hole decay, 
the classical evaporation is no longer  a good description. In such cases where the black hole mass is  near the Planck scale, 
 the burst model adopted by the \blackmax{} event generator~\cite{Dai:2007ki,Dai:2009by} is used to model the final part of the decay.

A search for microscopic black holes in a multijet final state is presented in Ref.~\cite{Khachatryan:2010wx}. In this analysis, 
events are selected  containing  two muons of the same charge.  This channel is expected to have low Standard Model backgrounds while
retaining good signal acceptance.
Isolated muons (i.e.\ muons with very little activity around them in the detector) can be 
produced directly from the black hole or from the decay of heavy particles such as $W$ or $Z$ bosons.
Muons from the semi-leptonic decays of heavy-flavour hadrons produced from the black hole can have several other particles nearby and can therefore
be non-isolated.
In order to maintain optimal acceptance for a possible signal, only one of the muons is required to be isolated
in this analysis, thereby typically increasing the acceptance in the signal region by 50\%.

The decay of the black hole to multiple high-$\pt$ objects is used to divide the observed events into background-rich 
and potentially signal-rich regions. This is done by using the number of high-$\pt$ charged particle tracks as the criterion to assign events to 
each region.  As will be quantified below, black hole events typically have a high number of tracks per event ($\ntrk$), while Standard Model processes 
have sharply falling track multiplicity distributions.
In the background-rich region, where only small signal contributions are expected, data and Monte Carlo simulations are used
to estimate the number of events after selections. This background estimate is validated by comparing to data.
The expected number of events from
Standard Model processes in the signal-rich region is then compared with the measured number, and a constraint on the contribution from
black hole decays is inferred.

The backgrounds from Standard Model processes are divided into two categories: processes where the two muons come from correlated decay chains and
processes that produce same-sign dimuons in uncorrelated decay chains. 
Same-sign dimuon events in correlated decay chains are produced primarily in the decays
of $\ttbar$ events and $\bbar$ events. In $\ttbar$ events, the most likely case is that the leading isolated muon arises from the decay of
a $\Wboson$-boson from one of the top-quarks, and the other muon of same charge comes from the semileptonic decay of a $b$-quark from the other
top-quark. In $\bbar$ events, the leading muon arises from the semileptonic decay of one $b$-quark, and the other muon from the sequential decay 
$b \rightarrow cX \rightarrow \mu X'$. Same-sign dimuons can also be produced due to $B^{0}\bar{B}^{0}$ mixing.
The backgrounds from $\ttbar$ and $\bbar$, and those from gauge boson pair production such as $WZ$ are estimated from Monte Carlo samples. 

Dimuon events in uncorrelated decay chains arise predominantly from the
$\Wboson+$jets process, where the leading isolated muon comes from $\Wboson$-boson decay and the other muon from a $\pi / K$ decay-in-flight, 
or the semi-leptonic decay of a $b$ or $c$ hadron in the remainder of the event. This background also has contributions from the $\Zboson +$jets process, and 
from low-$\pt$ dijet events.
The background from uncorrelated decay chains is estimated from data.
In the signal-rich
region, the dominant backgrounds come from  $\ttbar$ events and from muons produced in uncorrelated decays.

The rest of this paper is organised as follows. After a brief description of the ATLAS detector in Section~\ref{sec:atlas}, the data set and Monte Carlo samples
are described in Section\ref{sec:samples}. The event selection and the procedures to determine the backgrounds and their uncertainties are explained in Section~\ref{sec:evsel}
and~\ref{sec:bkgd} respectively. The results and 
their interpretation 
are discussed in Section~\ref{sec:result}.


\section{The ATLAS Detector}
\label{sec:atlas}

The ATLAS detector~\cite{atlas} covers nearly the entire solid angle\footnote{ATLAS uses a right-handed coordinate
system with its origin at the nominal interaction point (IP) in the centre of the detector and the $z$-axis coinciding
with the axis of the beam pipe. The $x$-axis points from the IP to the centre of the LHC ring, and the $y$-axis points
upward. Cylindrical coordinates ($r$,$\phi$) are used in the transverse plane, $\phi$ being the azimuthal angle
around the beam pipe, referred to the x-axis. The pseudorapidity is defined in terms of the polar angle $\theta$ (with respect to the z-axis)
as $\eta = -$ln tan$(\theta / 2)$.} 
around the collision point with layers of tracking detectors, calorimeters and muon chambers. The inner detector is immersed in 
a 2 T magnetic field along the $z$-axis and provides charged particle tracking in the range $|\eta|<2.5$. The silicon pixel
detector covers the vertex region and typically provides three measurements per track, followed by the silicon microstrip
tracker (SCT) which provides measurements from eight strip layers. The silicon detectors are complemented by the
transition radiation tracker (TRT) which provides more than 30 straw-tube measurements per track and improves the 
momentum resolution.

The calorimeter system covers the pseudorapidity range $|\eta| < 4.9$. Lead-liquid argon (LAr) electromagnetic sampling calorimeters
cover the range $|\eta| < 3.2$, with an additional thin LAr presampler covering $|\eta| < 1.8$ to correct for energy loss in
material upstream of the calorimeters. Hadronic calorimetry is provided by a scintillator-tile calorimeter 
over $|\eta| < 1.7$  and two copper/LAr endcap calorimeters over $1.75< |\eta| < 3.2$. The solid
angle coverage is completed with forward copper/LAr and tungsten/LAr calorimeters for electromagnetic and hadronic measurements 
respectively up to $|\eta| < 4.9$.

The muon spectrometer consists of separate trigger and high-precision tracking chambers which measure the deflection of muon tracks
in a magnetic field with a bending integral of approximately 2 to 8 Tm. The magnetic field is generated by three
superconducting air-core toroid magnet systems. The tracking chambers cover the region $|\eta|<2.7$ with three layers of monitored drift tubes
 and cathode strip chambers in the innermost region of the endcap muon spectrometer.
The muon trigger system covers the range $|\eta|<2.4$ with resistive plate chambers in the barrel, and thin gap chambers in the
endcap regions.


\section{Data and Monte Carlo Samples}
\label{sec:samples}

The data used in this analysis were collected between March and July 2011 at the LHC  operating at a centre
of mass energy of 7 $\tev$. The total integrated luminosity after detector and data-quality requirements is
1.3 $\ifb$, with an uncertainty of 3.7\%~\cite{Aad:2011dr,lumi}. The data were recorded with a single muon trigger with the threshold
at $20\gev$ on the muon's transverse momentum. The muon trigger efficiency reaches the plateau regime for transverse momenta above $25 \gev$. The
plateau efficiency is 75\% in the barrel and 88\% in the endcap for muons reconstructed offline.
In this analysis it is required that at least one of the selected muons with $\pt$ above $20\gev$ matches the trigger criteria.
During the considered data-taking period, the LHC configuration was such that the mean number of  primary
proton-proton interactions per bunch crossing was close to 6. The effect of this ``pile-up'' is taken into account 
in the analysis.

Several Monte Carlo samples are used both for signal modelling and background estimation.
 These samples are processed with the 
ATLAS full detector  simulation~\cite{bib:atlas-sim} which is based on the 
{\sc GEANT4} toolkit~\cite{bib:g4}. The simulated events are then reconstructed with the same software chain as the data. 
The effect of pile-up is modelled by overlaying simulated
minimum bias events onto the original hard-scattering event. Monte Carlo events are then re-weighted so that the reconstructed
vertex multiplicity distribution agrees with the data.

Background Monte Carlo samples are generated for $\ttbar$, as well as for $\bbar$ and $\ccbar$ processes. The latter are 
considered together in the following and referred to as  $b/c$ for simplicity.  The $\ttbar$ events are generated with 
\mcnlo~\cite{mcnlo1,mcnlo2} with an assumed top-quark mass of 172.5$\gev$, and with the next-to-leading order CTEQ66~\cite{cteq} parton distribution 
function (PDF) set. Fragmentation  and hadronisation of the events is done with \herwig~\cite{herwig} using \jimmy~\cite{jimmy} for 
the underlying event model. The $b/c$ Monte Carlo sample is generated and hadronised with \pythia~\cite{pythia} 
using the ATLAS AMBT1 tune~\cite{Aad:2010ir}.
It is produced with a filter at the generator level requiring two muons with $\pt > 10 \gev$ each.  
The diboson samples ($WZ$ and $ZZ$) are generated with \herwig. They are filtered to have at least one electron or muon
with $\pt > 10 \gev$. The single top background in the $Wt$ channel is estimated with \acermc~\cite{acermc}. The single top
backgrounds in the $t-$channel and $s-$channel are included in the background estimate derived from data.
The $b/c$, diboson and single top samples are all generated using the MRST2007 PDF~\cite{mrst}.
Samples of $W+$jet and $Z+$jet events produced using Alpgen~\cite{alpgen} are also used for cross-checks.

Signal Monte Carlo samples are generated using  \blackmax~2.01 and hadronised with \pythia{} using the ATLAS AMBT1 tune. The samples
are produced with the CTEQ66 PDF set with the mass of the black hole used as the QCD scale. 
For the signal samples, $M_{D}$ is varied between 0.5$\tev$ and 2.5$\tev$ and $\MTH$ is varied between 3$\tev$ and 5$\tev$. 
In each case, samples are generated with 2, 4 and 6 extra dimensions.


\section{Event Selection}
\label{sec:evsel}

Events passing the single muon trigger are required to have at least one primary vertex with at least five associated tracks with $\ptTrack > 400 \mev$.
If the event has multiple primary vertices, the vertex with the largest $\sum (\ptTrack )^2$ is identified as the ``hard-scattering vertex". 

Tracks found in the inner detector (ID) are selected using the following criteria:
\begin{eqnarray*}
&\ptTrack >1\,\GeV, \ N_{\rm pixel} \ge 1, \ N_{\rm SCT} \ge 6, & \\
&|\eta| < 2.4, \ |d_0| < 1.5\,{\rm mm}, \ |z_0\sin\theta| < 1.5\,{\rm mm},& 
\end{eqnarray*}
where $N_{\rm pixel}$ and $N_{\rm SCT}$ are the number of hits\footnote{A hit is a signal above threshold in a particular detector element.} from the
pixel and the SCT detectors, respectively, that are associated with the 
track, and
$d_0$ and $z_0$ are the transverse and longitudinal impact parameters 
measured with 
respect to the hard-scattering vertex.  Muon candidates are reconstructed from tracks measured in the muon spectrometer (MS). The MS tracks are then matched with ID tracks
using a procedure that takes material effects into account. The parameters for the resulting matched
muon candidates are obtained by a statistical combination of the measurements in the MS and the ID.
 
At least two muons  passing these selections are required in each event. Both must come from the hard-scattering vertex and satisfy $|\eta|<2.4$.
 The muon with the highest transverse momentum is required
to have $\pt > 25 \gev$. This leading muon is also required to be isolated by requiring that the sum of transverse momenta of ID tracks in
a cone in $\eta-\phi$ space of radius $\Delta R= 0.2$ around the muon is less than 0.2$\times \pt$ of the muon. The muon with the next highest transverse momentum is
required to have $\pt > 15 \gev$ and the same charge as the leading muon.  No 
isolation requirement is made on this second muon.  The two muons are required to satisfy
$\Delta R > 0.2$ to explicitly ensure 
that the isolated muon is not close to the second muon.
 The leading muon is required to have small impact parameter
significance by imposing $|d_0/\sigma(d_0)| < 3$. The impact parameter is calculated  with respect to the hard-scattering  vertex in the event. 

The track multiplicity is constructed by counting the number of ID tracks associated to the hard-scattering vertex
which satisfy $\pt > 10 \gev$ and $|\eta|<2.4$.  By definition, the track count
includes the two muon candidates. A signal-rich region is defined by selecting events with at least ten such tracks, while events with less than
ten tracks are used to validate the prediction of the expected backgrounds.

All selections except the trigger are applied to the  Monte Carlo events.
To account for the trigger efficiency, the Monte Carlo events are weighted with the efficiency measured from data, while the differences in muon
reconstruction and identification between data and simulation are accounted for by applying  $\pt$ and $\eta$ dependent 
scale factors~\cite{muid1,muid2} to the Monte Carlo events
when calculating the acceptance. This is important when the sub-leading muon provides the trigger as the trigger efficiency varies
with $\pt$ in the region between 20 and 25 \gev.

The tracking efficiency in data is well reproduced by the Monte Carlo simulation~\cite{minbias}. This is confirmed by additional studies of tracking
performance in a dense environment~\cite{Aad:2011sc}. No corrections to the Monte Carlo are therefore applied.


\section{Background Estimation}
\label{sec:bkgd}

The two components of the background from correlated and uncorrelated particle decays are determined using a mixture of Monte Carlo simulation and 
techniques using data. 


\subsection{Correlated Background Estimates}
\label{sec:mcbkg}

 The correlated background arises from processes such as $\ttbar$ production where, for example, the isolated muon comes from 
top decay ($t\to bW \to b \mu\nu$) and the other (non isolated)  from the antitop decay ($\overline{t} \to W \overline{b} \to  W \mu \nu \overline{c}$).
The background from $\ttbar$ production is estimated from Monte Carlo simulation. The approximate next-to-next-to-leading-order production cross section of 
165~pb~\cite{Moch:2008qy,Langenfeld:2009tc,topxsec} is used to normalise the Monte Carlo prediction.  This cross section is in agreement with the measurement of the $\ttbar$
cross section at ATLAS~\cite{topxsec1fb}.
 The sources of systematic uncertainty on the $\ttbar$ background described in Ref.~\cite{Aad:2010ey} are considered and the uncertainty from each source
is shown in Table~\ref{tab:syst}.
The sources considered are the
choice of generator, the amount of initial and final state radiation (ISR/FSR), the top-quark mass, and the theoretical uncertainty on the predicted production
cross section. The largest contribution to the uncertainty is
9.6\% on the cross section which arises from variations in the renormalisation and factorisation scales (5.6\%) and the PDF uncertainty (4\%).
The uncertainty due to the choice of generator is evaluated 
by comparing the predictions of \mcnlo{} with those of \powheg~\cite{powheg} interfaced to \pythia. The \powheg{} samples are generated
using the  MRST2007 PDF set. The uncertainty due to the top-quark
mass is obtained by generating $\ttbar$ samples with top mass $\pm 2.5 \gev$ from the nominal choice of 172.5 \gev. The ISR/FSR uncertainty is determined
by using the \acermc{} generator interfaced to \pythia, and by varying the ISR and FSR $\Lambda_{QCD}$, and the ISR and FSR cutoff.
There is also an additional
2.6\% uncertainty on the $\ttbar$ estimate from trigger weight and muon reconstruction  efficiency scale factors.

\begin{table}
  \begin{center}
    \begin{tabular}{ccccc}
\hline \hline
Source & 	Muon$+$fake (\%)& 	$\ttbar$ (\%)& 	$b/c$ (\%)& Signal (\%)\\ 
\hline
Measurement statistics& 4.1&            &               &\\
Subtraction of& 	20 & 		& 		&\\
$\ttbar+b/c+Wt+$diboson& &            &               &\\ \hline
ISR/FSR& 	& 		7.1 & 		&\\
$t$-quark mass& & 		4.5 & 	 	&\\
Cross section& 	& 		${}^{+7}_{-9.6}$&   &\\
Monte Carlo Generator& 	& 		5.1 & 		&\\
Luminosity&	&		3.7 &		 & 3.7\\
$\mu$ reco/trig& &		2.6 &		 & 1.5\\
PDF(Acceptance)& & & & 3.0\\
Rescaled Truth Acceptance& & & & 14.3\\ \hline
Ratio (nominal/inverted)& 	& 		& 		8.5 &\\
Extrapolation to $\ntrk$ $\ge 10$& & & 100& \\
\hline
Total uncertainty& 20.4& 14.4& 100.4 & 15.1\\
\hline
\end{tabular}
  \caption[Systematic Uncertainties]{Systematic uncertainties in percent on the background and signal estimates in the signal region from various sources. 
$\mu$ reco/trig stands for the uncertainty due to trigger efficiency and
muon reconstruction efficiency scale factors applied to the Monte Carlo events. A blank entry indicates that the particular systematic uncertainty does
not apply to that particular background.
}
  \label{tab:syst}
  \end{center}
\end{table}

The background from $b/c$ production is estimated in two steps. In the first step, the background is determined in the 
$\ntrk$ $<10$ (background) region using a heavy-flavour enriched data sample to normalise the Monte Carlo prediction. In the second step, the estimated background is 
extrapolated from $\ntrk$  $<10$ to $\ntrk$ $\ge 10$ using Monte Carlo. 

To estimate $b/c$ production in the background region, a heavy-flavour rich sample is selected by inverting the isolation and impact 
parameter significance requirements on the leading muon. 
This yields 6480 events. The  $b/c$ Monte Carlo sample is used to measure
the ratio of events passing the nominal muon selection to those passing the inverted selection. The ratio is $0.33 \pm 0.03$ where the uncertainty comes from
the limited size of the Monte Carlo sample. Applying this
ratio to the heavy-flavour rich sample in data gives the  $b/c$ estimate in the background region. The shapes of the
kinematic distributions for the $b/c$ background, such as $\pt$ of the muons are also obtained from the heavy-flavour
rich sample.

The $\ntrk$ distribution in the Monte Carlo is then fit with an exponential to determine the fraction of events
with $\ntrk \ge 10$. 
The method is validated by varying the fit range, testing the extrapolation procedure in the $\ttbar$ Monte Carlo, as well as by
 relaxing the $\pt$
requirements on the muons to enhance the statistics of the $b/c$ Monte Carlo. Based on these studies, a 100\% systematic
uncertainty due to the extrapolation is assigned to the $b/c$ background in the signal region. 


The backgrounds from diboson  ($WZ, ZZ$) and  single-top processes 
 are estimated from the corresponding Monte Carlo samples and are found to be negligible.
\subsection{Uncorrelated Background Estimate}

\label{sec:datbkg}

The uncorrelated background arises when the second muon is not a true muon (fake), or 
is a muon from $K$ or $\pi$ decay, or from events where there 
is no correlation between the production mechanisms of the two muons.
 The background from uncorrelated decays  is estimated by first measuring the probability for a track to be reconstructed as a muon
in a control sample from data. This `fake' probability is then applied to data events with one muon and one or more tracks
to obtain a prediction for $\mu +$fake dimuon events. 

The control sample consists of $\Wboson$-boson + track events. Events are selected with
at least one isolated muon with $\pt > 25 \gev$ and missing transverse momentum ($\met$) satisfying $25 \gev < \met < 80 \gev$.
$\met$ is constructed from the
sum of all cells contained in calorimeter clusters and is corrected for the presence of muons in the event.
 The transverse mass
calculated from the muon and the $\met$ 
is required to be between 50$\gev$ and 120$\gev$. These events are also required to have at least one track in addition to the muon,
with $\pt > 15 \gev$ and the same charge as the muon. If an event has more than one such track,
then all tracks are considered for the measurement. The events are also required to have less than
ten tracks to remove possible  signal contributions.  

A subset of the $\Wboson$-boson + track control sample is then selected by requiring an additional muon passing the analysis selection criteria
 with $\pt > 15 \gev$ and of the same charge as the first selected muon.  
Using this subset, the fraction of events where a second muon is present is determined directly from data.
This fraction contains contributions from fakes, $K$ or $\pi$ decay, and heavy-flavour decays.  To
avoid double-counting, the contributions from the correlated decays from $\ttbar$, $b/c$, and diboson processes
 are estimated from Monte Carlo (as described) and subtracted. 
The per-track rate is measured in three $\pt$ bins; for $\pt < 20 \gev$ the rate is $(4.4 \pm 0.2) \times 10^{-3}$, for $20 < \pt < 60 \gev$ the rate is
$(3.7 \pm 0.1) \times 10^{-3}$, and for $\pt > 60 \gev$ the rate is $(3.7 \pm 2) \times 10^{-3}$.
This rate is applied to all events in data with one muon and at least one track of 
the same charge with $\pt > 15 \gev$. If more than one track is found, then each track is considered in calculating a total probability for the event to 
be reconstructed as a dimuon event. The uncertainty on the background estimate from the nominal fake rate measurement due to measurement statistics
is 4.1\%.

To determine the systematic uncertainty, shown in Table~\ref{tab:syst}, the amount of subtracted $\ttbar + b/c + Wt +$diboson background is varied up
and down by $1\sigma$. The corresponding variation in the fake estimate is 20\% which is taken as the systematic uncertainty on this background.

This method is verified by using the $W+$jet and single-top Monte Carlo samples as pseudo-data to measure the rate and then make a prediction. 
Similar studies on fake muon probability, with different selection criteria, are reported in Ref.~\cite{Aad:2011rr} and show consistent results.

The background estimation is tested in events with the same selections as the signal region except the track multiplicity which is required to be $\ntrk < 10$.  
The prediction from the Standard Model is shown in Table~\ref{tab:crlontrk}, along with the number of observed events in data in the background region.
The contribution from the signal in the background region has been checked to be less than 0.1\% of backgrounds for various choices of the signal parameters.
The event rates observed in the background region agree with the prediction within the uncertainties.

\begin{table}
  \begin{center}
    \begin{tabular}{cc}
      \hline \hline
	Process & Events\\
      \hline
      $b/c$& 2120 $\pm$  30(stat) $\pm$ 200(syst) \\
    $\ttbar$& 750 $\pm$ 100(syst) $\pm$ 30(lumi) \\
    $\mu +$fake& 1300 $\pm$ 2(stat) $\pm$ 260(syst)\\
        $Wt$& 53 $\pm$ 2(syst)\\
        $WZ+ZZ$& 36 $\pm$ 1(syst)\\ \hline
    Predicted& 4270 $\pm$ 30(stat) $\pm$ 340(syst)\\
    Observed& 3775 \\
      \hline	
    \end{tabular}
  \caption[Low NTracks control region]{ Numbers of expected and observed events in the background-rich control region with $\ntrk < 10$. 
Only the uncertainties due to the limited size of the Monte Carlo samples are included for the diboson ($WZ+ZZ$) and single-top ($Wt$) backgrounds.
}
  \label{tab:crlontrk}
  \end{center}
\end{table}

\begin{figure}
  \centering
     \includegraphics[width=0.49\textwidth]{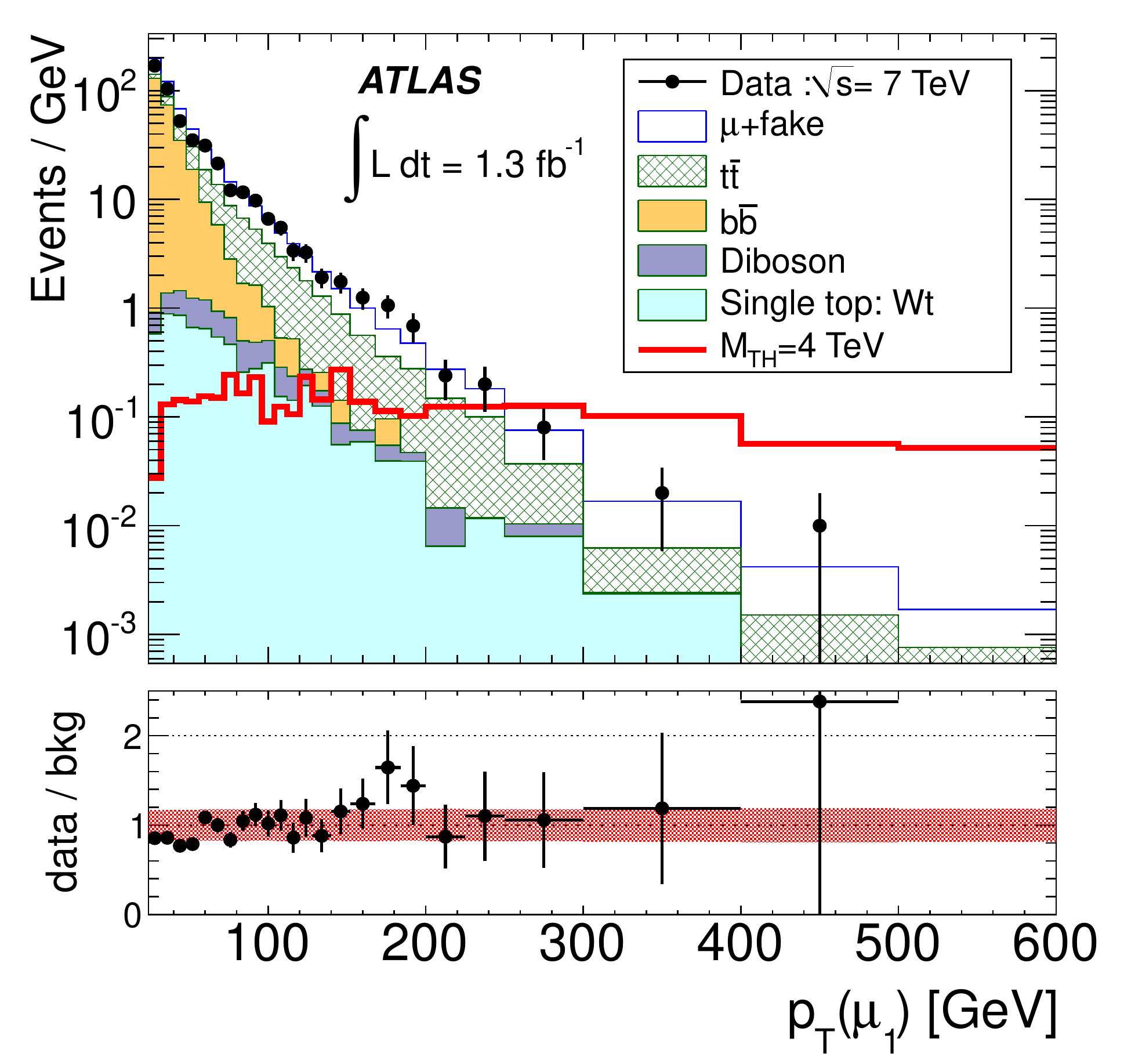}
     \includegraphics[width=0.49\textwidth]{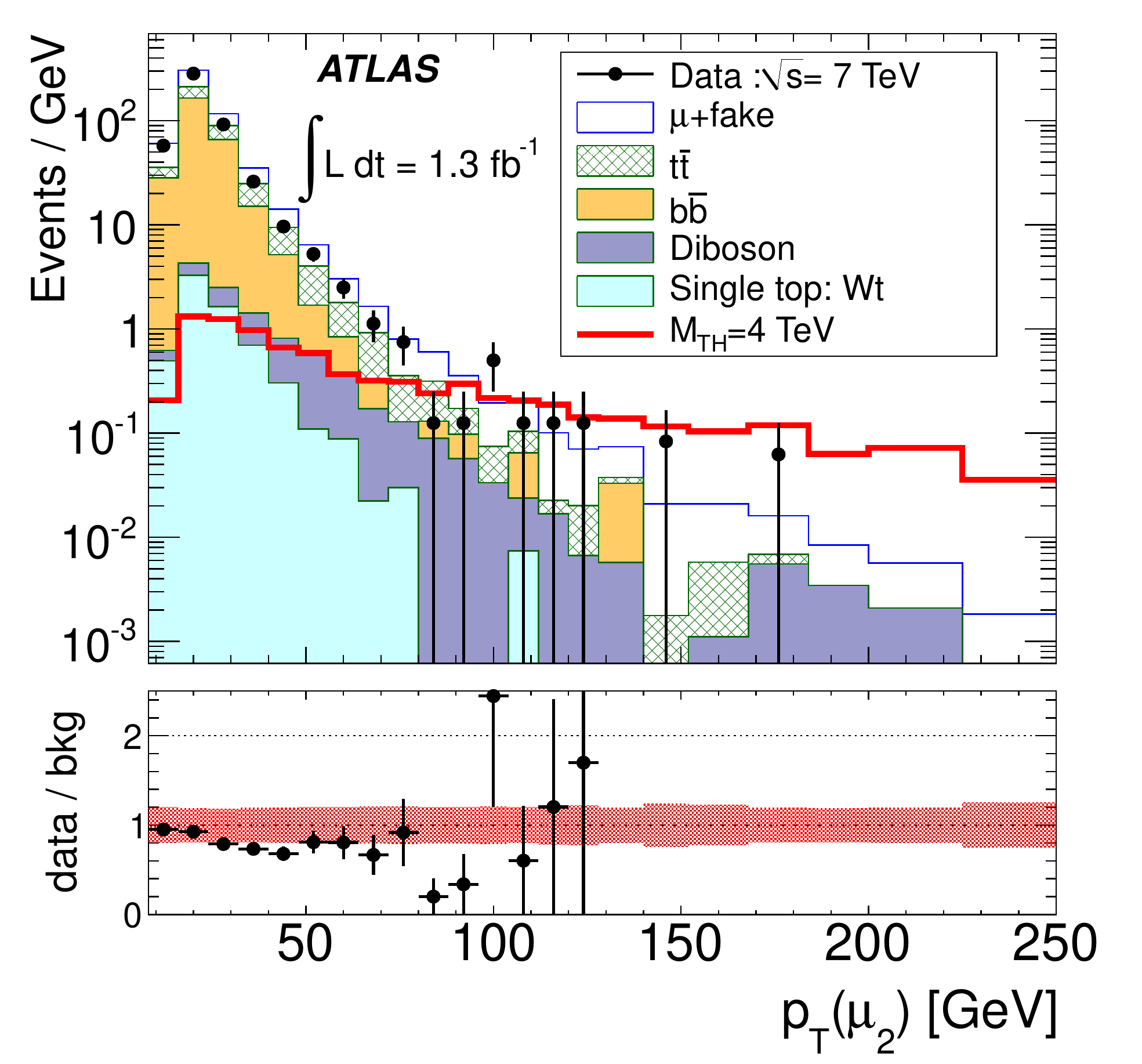}
  \caption{The leading (left) and sub-leading (right) muon $\pt$ distributions for same-sign dimuon events before the $\ntrk$ cut. The background histograms are stacked. 
The signal expectation for
a non-rotating black hole model with parameters $\MD = 800 \gev$, $\MTH=4\tev$, and six extra dimensions is overlaid for illustrative purposes. The bottom panels
show the ratio of data to the expected background (points) and the total systematic uncertainty on the background (shaded area).
}
  \label{fig:crlspt}
\end{figure}


\section{Results and Interpretation}
\label{sec:result}

\begin{figure}
  \centering
    \includegraphics[width=0.7\textwidth]{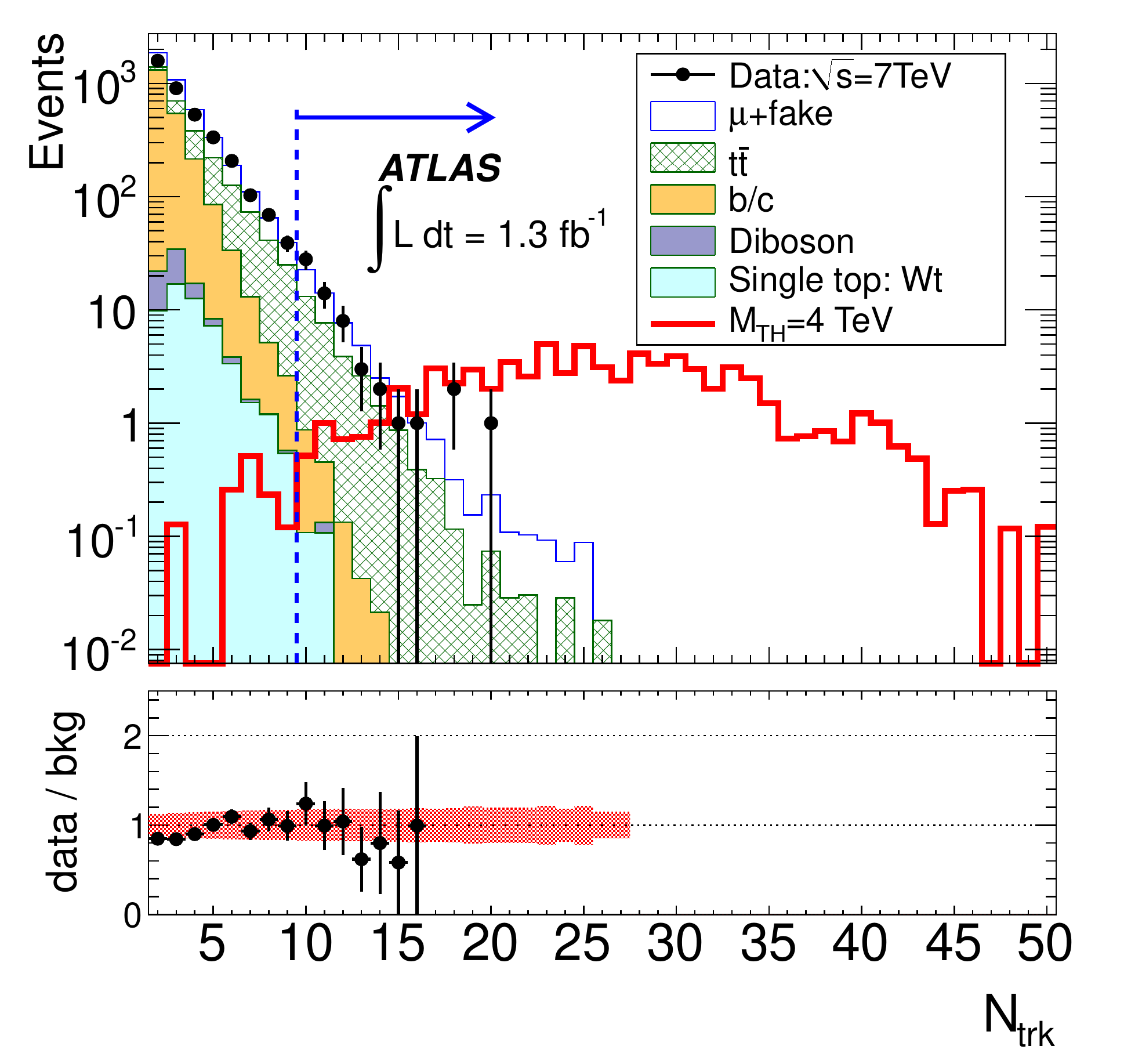}
  \caption{The track multiplicity distribution for same-sign dimuon events. The region with  $\ntrk \ge 10$ is selected as the signal region.
The background histograms are stacked. The signal expectation for
a non-rotating black hole model with parameters $\MD = 800 \gev$, $\MTH=4 \tev$, and six extra dimensions is overlaid for illustrative purposes.
The bottom panel
shows the ratio of data to the expected background (points) and the total systematic uncertainty on the background (shaded area).}
    \label{fig:crlsntrk}
\end{figure}

Figures~\ref{fig:crlspt} and \ref{fig:crlsntrk} show the $\pt$ distributions of both muons and the track multiplicity in all 
same-sign dimuon events respectively before applying the $\ntrk$ requirement.
The prediction for a sample signal model for non-rotating black holes with $\MD = 800 \gev$,  $\MTH=4 \tev$, and six extra dimensions is also shown.
Good agreement is observed between the measured distributions and the background expectations.
 As shown in Figure~\ref{fig:crlsntrk}, the backgrounds peak at low
values of the track multiplicity while a possible signal has a higher number of tracks.   
 Table~\ref{tab:res} shows the expected and observed numbers of same-sign dimuon events in the signal region. 
No excess over the Standard Model predictions is observed in the data. 

The background in the signal region is dominated by the
$t\overline{t}$ and by the uncorrelated decays from $W+$jet events.
The relative contributions of the various backgrounds are different in the background-rich (Table~\ref{tab:crlontrk}) and signal-rich 
(Table~\ref{tab:res}) regions. In particular the $b/c$ contribution falls more rapidly with increasing $\ntrk$ than the other backgrounds
and is very small in the signal-rich region. By removing the isolation requirement on the leading muon, the distribution is dominated by
$b/c$ background and the Monte Carlo simulation agrees with data giving confidence in the $b/c$ prediction.

\begin{table}[htbp]
  \begin{center}
    \begin{tabular}{cc}
      \hline \hline
	Process & Events\\
      \hline
	$b/c$& 0.77 $\pm$ 0.77(syst)\\
	$\ttbar$& 29.2 $\pm$ 4.1(syst) $\pm$ 1.1(lumi) \\
	$\mu +$fake&  25.6$\pm$ 0.3(stat) $\pm$ 5.2(syst) \\ 
        Other backgrounds& 0.25 $\pm$ 0.11(syst) \\ \hline
	Predicted& 55.8 $\pm$ 0.3(stat) $\pm$ 6.7(syst) $\pm$ 1.1(lumi)\\
	Observed& 60\\ \hline
	Signal $M_{TH}=4 \tev$& 72.1 $\pm$ 4.5(syst)\\
      \hline	
    \end{tabular}
  \caption[Signal region]{ Number of expected and observed events in the signal region, like-sign dimuon events with 
$ \ntrk \ge 10$. The other backgrounds are from diboson and single-top processes.
The signal expectation for a non-rotating black hole model with $\MD = 800 \gev$, $M_{TH}=4 \tev$, and six extra dimensions 
is also shown.
}
  \label{tab:res}
  \end{center}
\end{table}


Using the number of events observed in data and the background expectations, upper limits are set on $\sigma \times BR \times A$, where
$\sigma$ is the cross section, $BR$ the branching ratio to dimuons, and $A$ the acceptance of non Standard Model contributions in this final state
in the signal region. 
The CLs method~\cite{cls} is used to derive these limits assuming  Gaussian uncertainties on the predicted background and signal, and
Poissonian fluctuations on the observed number of events.  The observed 95\% confidence level upper limit on
$\sigma\times BR \times A$ is 0.018~pb. This result is compatible with the expected limit of 0.016~pb, which is determined 
from pseudo-experiments using simulation. 
The $1\sigma$ and $2\sigma$ ranges on the expected limit are from 0.012 to 0.022 pb and from 0.008 to 0.029 pb respectively. The $BR \times A$ for the
signal model shown in Table~\ref{tab:res} is 3\%, and typically varies between 1\% and 6\% for the signal models considered here.

Limits on the reduced Planck mass ($\MD$) and the minimum mass of the black hole ($\MTH$) for several models are set using the \blackmax{} generator and the CTEQ66 PDF. 
The signal yield is affected by the PDF choice due to two distinct effects: the change in the production cross section
and the change in signal acceptance. 
The signal cross section obtained from MRST2007 is typically 40\% to 50\% higher than that from CTEQ66 for 
$M_D=1\tev$, $M_{TH}=4 \tev$. This difference is somewhat larger than the uncertainty on the cross section from the CTEQ66 PDF error sets. 
At the large values of $\MTH$ near the quoted limits, 
the invariant mass of the incoming partons is large and the PDFs are therefore used in a range of parton momentum fraction 
$x$ where they are not well constrained. The theoretical uncertainty on the production cross section is potentially very large.
For these reasons, no theoretical uncertainty on the signal cross section is assigned, that is, 
the exclusion limits are set  for the exact benchmark models as implemented in the \blackmax{} generator: using CTEQ66 rather than MRST2007 gives a more
conservative limit. The cross section for the signal point shown in Table~\ref{tab:res} is 2.1 pb. 
The uncertainty on the signal acceptance from the choice of PDF is estimated to be 3\% by using
the 44 error sets of the CTEQ66 PDF and is a small contribution to the overall uncertainty.

The observed results are used to obtain exclusion contours in the plane of \MD{} and \MTH{}.
For a large number of points in the
$(\MD,\MTH)$ plane, the signal acceptance is measured using kinematic properties obtained from the event generator (truth).
This truth level acceptance is compared to
the acceptance from full detector simulation for a smaller set of points which are representative of the model parameters probed in
this analysis. 
To account for the difference in acceptances, the truth
level acceptance is scaled by a constant factor of $0.7\pm0.1$ which is determined by comparing truth to fully simulated points.  
Therefore the
uncertainty on the signal prediction consists of the following components:  the uncertainty due to rescaling of truth acceptance, the uncertainty
on the luminosity of the data sample, the uncertainty on acceptance due to the PDF, the experimental uncertainty on acceptance due to muon trigger and identification 
efficiencies and a statistical uncertainty due to the finite Monte Carlo samples (see Table~\ref{tab:syst}).

Figure~\ref{fig:limsum} shows the expected and observed exclusion contours for rotating and 
non-rotating black holes for 2 and 6 extra dimensions. The non-smoothness of the exclusion 
contours reflects the discrete nature of the Monte Carlo grid in the $(\MD,\MTH)$ plane and 
the finite Monte Carlo statistics at the generated points. Lines of constant slope ($\MTH/\MD$) of 3, 4 and 5
are also shown in the figure. The semi-classical approximations used for black hole production 
and decay are expected to be valid only for large slopes. It can be seen
that if this ratio is greater than three, the limit on $\MTH$ is larger than half
the centre-of-mass energy. 

In view of the rapidly falling PDF's in this region, further significant  improvements on these limits 
are not expected until the LHC energy is increased. For example, moving from $\MTH=4.7\tev$ to $\MTH=5\tev$ changes the signal 
cross section from 0.24 pb to 0.06 pb (for non-rotating black holes in models with $\MD=500\gev$ and six extra dimensions).
It is also worth noting that the exclusion contours are dependent on the model considered, and this analysis is not expected to be
sensitive to black hole models with decays to low multiplicity final states such as quantum black holes~\cite{Meade:2007sz}.


\newlength{\mdmthplotwidth}
\setlength{\mdmthplotwidth}{0.4\textwidth}

\begin{figure}
  \centering
    \includegraphics[width=0.49\textwidth]{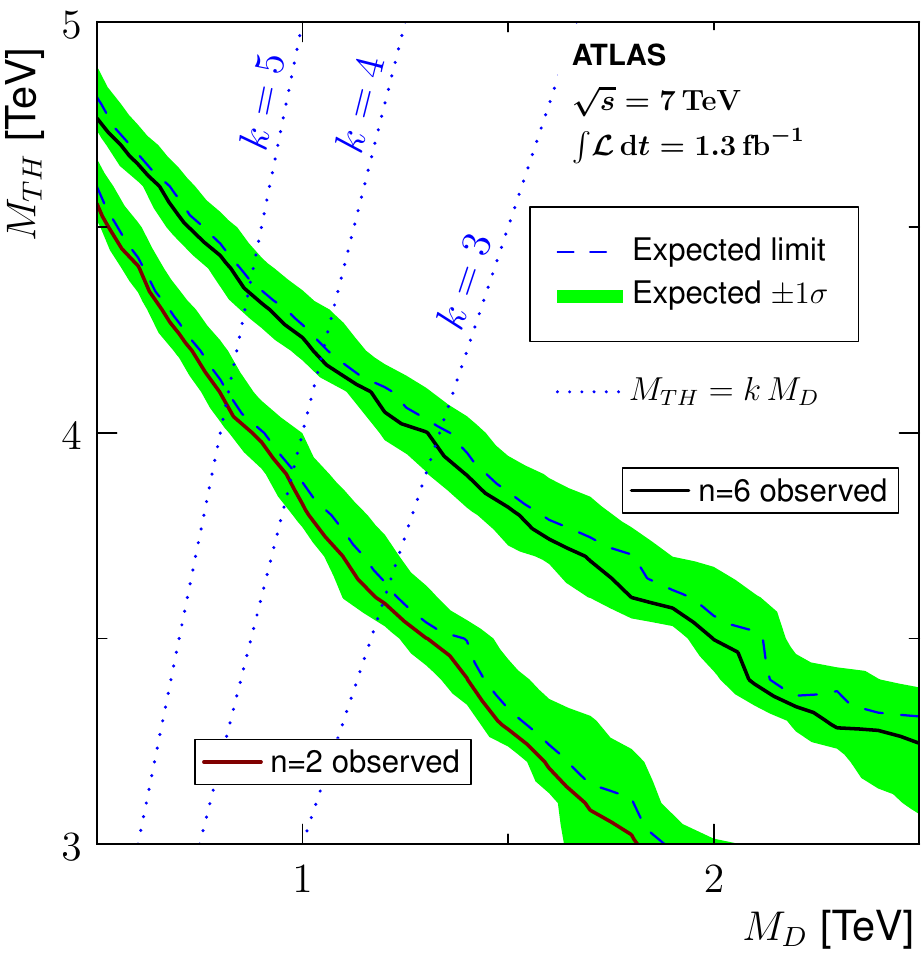}
    \includegraphics[width=0.49\textwidth]{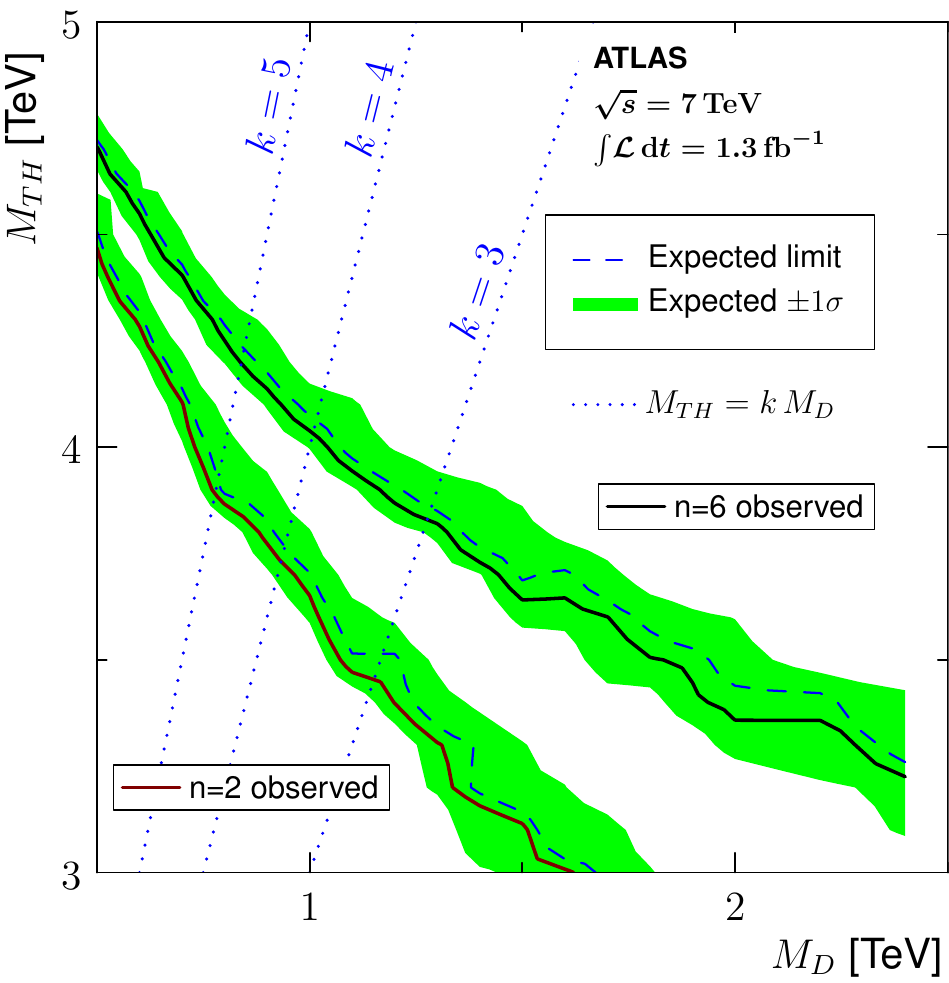}
  \caption{95\% confidence level exclusion contours for non-rotating (left) and rotating (right) black holes in models with two and six extra dimensions.
The dashed lines show the expected exclusion contour with the
$1\sigma$ uncertainty shown as a band.  The solid lines show the observed exclusion contour. The regions below the contour
are excluded by this analysis. The dotted lines show lines of constant slope equal to 3, 4, and 5. Only slopes much larger than 1 correspond to physical models.}
    \label{fig:limsum}
\end{figure}

In summary, a search for extra dimensions in the same-sign dimuon final state has been performed using 1.3 $\ifb$ of data recorded with the ATLAS detector
in $7 \tev$ proton-proton collisions at the LHC. No excess of events over the Standard Model prediction is observed and
exclusion contours are obtained in the plane of the reduced Planck scale $\MD$ and the threshold $\MTH$ for black hole production.
A model independent limit of 0.018 pb on any new physics contribution in the signal region with the described selection is set.

\section{Acknowledgements}

We thank CERN for the very successful operation of the LHC, as well as the
support staff from our institutions without whom ATLAS could not be
operated efficiently.

We acknowledge the support of ANPCyT, Argentina; YerPhI, Armenia; ARC,
Australia; BMWF, Austria; ANAS, Azerbaijan; SSTC, Belarus; CNPq and FAPESP,
Brazil; NSERC, NRC and CFI, Canada; CERN; CONICYT, Chile; CAS, MOST and
NSFC, China; COLCIENCIAS, Colombia; MSMT CR, MPO CR and VSC CR, Czech
Republic; DNRF, DNSRC and Lundbeck Foundation, Denmark; ARTEMIS, European
Union; IN2P3-CNRS, CEA-DSM/IRFU, France; GNAS, Georgia; BMBF, DFG, HGF, MPG
and AvH Foundation, Germany; GSRT, Greece; ISF, MINERVA, GIF, DIP and
Benoziyo Center, Israel; INFN, Italy; MEXT and JSPS, Japan; CNRST, Morocco;
FOM and NWO, Netherlands; RCN, Norway; MNiSW, Poland; GRICES and FCT,
Portugal; MERYS (MECTS), Romania; MES of Russia and ROSATOM, Russian
Federation; JINR; MSTD, Serbia; MSSR, Slovakia; ARRS and MVZT, Slovenia;
DST/NRF, South Africa; MICINN, Spain; SRC and Wallenberg Foundation,
Sweden; SER, SNSF and Cantons of Bern and Geneva, Switzerland; NSC, Taiwan;
TAEK, Turkey; STFC, the Royal Society and Leverhulme Trust, United Kingdom;
DOE and NSF, United States of America.

The crucial computing support from all WLCG partners is acknowledged
gratefully, in particular from CERN and the ATLAS Tier-1 facilities at
TRIUMF (Canada), NDGF (Denmark, Norway, Sweden), CC-IN2P3 (France),
KIT/GridKA (Germany), INFN-CNAF (Italy), NL-T1 (Netherlands), PIC (Spain),
ASGC (Taiwan), RAL (UK) and BNL (USA) and in the Tier-2 facilities
worldwide.


\bibliographystyle{model1-num-names-notitle}
\bibliography{conf}

\begin{thebibliography}{41}
\expandafter\ifx\csname natexlab\endcsname\relax\def\natexlab#1{#1}\fi
\providecommand{\bibinfo}[2]{#2}
\ifx\xfnm\relax \def\xfnm[#1]{\unskip,\space#1}\fi
\bibitem[{Arkani-Hamed et~al.(1998)Arkani-Hamed, Dimopoulos, and
  Dvali}]{ArkaniHamed:1998rs}
\bibinfo{author}{N.~Arkani-Hamed}, \bibinfo{author}{S.~Dimopoulos},
  \bibinfo{author}{G.~R. Dvali},
\newblock \bibinfo{journal}{Phys. Lett.} \bibinfo{volume}{B429}
  (\bibinfo{year}{1998}) \bibinfo{pages}{263--272}.
\bibitem[{Antoniadis et~al.(1998)Antoniadis, Arkani-Hamed, Dimopoulos, and
  Dvali}]{Antoniadis:1998ig}
\bibinfo{author}{I.~Antoniadis}, \bibinfo{author}{N.~Arkani-Hamed},
  \bibinfo{author}{S.~Dimopoulos}, \bibinfo{author}{G.~R. Dvali},
\newblock \bibinfo{journal}{Phys. Lett.} \bibinfo{volume}{B436}
  (\bibinfo{year}{1998}) \bibinfo{pages}{257--263}.
\bibitem[{Arkani-Hamed et~al.(1999)Arkani-Hamed, Dimopoulos, and
  Dvali}]{ArkaniHamed:1998nn}
\bibinfo{author}{N.~Arkani-Hamed}, \bibinfo{author}{S.~Dimopoulos},
  \bibinfo{author}{G.~R. Dvali},
\newblock \bibinfo{journal}{Phys. Rev.} \bibinfo{volume}{D59}
  (\bibinfo{year}{1999}) \bibinfo{pages}{086004}.
\bibitem[{Argyres et~al.(1998)Argyres, Dimopoulos, and
  March-Russell}]{Argyres:1998qn}
\bibinfo{author}{P.~C. Argyres}, \bibinfo{author}{S.~Dimopoulos},
  \bibinfo{author}{J.~March-Russell},
\newblock \bibinfo{journal}{Phys.Lett.} \bibinfo{volume}{B441}
  (\bibinfo{year}{1998}) \bibinfo{pages}{96--104}.
\bibitem[{Banks and Fischler(1999)}]{Banks:1999gd}
\bibinfo{author}{T.~Banks}, \bibinfo{author}{W.~Fischler}
  (\bibinfo{year}{1999}).
\bibitem[{Dimopoulos and Landsberg(2001)}]{Dimopoulos:2001hw}
\bibinfo{author}{S.~Dimopoulos}, \bibinfo{author}{G.~L. Landsberg},
\newblock \bibinfo{journal}{Phys. Rev. Lett.} \bibinfo{volume}{87}
  (\bibinfo{year}{2001}) \bibinfo{pages}{161602}.
\bibitem[{Giddings and Thomas(2002)}]{Giddings:2001bu}
\bibinfo{author}{S.~B. Giddings}, \bibinfo{author}{S.~D. Thomas},
\newblock \bibinfo{journal}{Phys. Rev.} \bibinfo{volume}{D65}
  (\bibinfo{year}{2002}) \bibinfo{pages}{056010}.
\bibitem[{Kanti(2004)}]{Kanti:2004nr}
\bibinfo{author}{P.~Kanti},
\newblock \bibinfo{journal}{Int. J. Mod. Phys.} \bibinfo{volume}{A19}
  (\bibinfo{year}{2004}) \bibinfo{pages}{4899--4951}.
\bibitem[{Dai et~al.(2008)}]{Dai:2007ki}
\bibinfo{author}{D.~C. Dai}, et~al.,
\newblock \bibinfo{journal}{Phys. Rev.} \bibinfo{volume}{D77}
  (\bibinfo{year}{2008}) \bibinfo{pages}{076007}.
\bibitem[{Hawking(1975)}]{Hawking:1974sw}
\bibinfo{author}{S.~Hawking},
\newblock \bibinfo{journal}{Commun. Math. Phys.} \bibinfo{volume}{43}
  (\bibinfo{year}{1975}) \bibinfo{pages}{199--220}.
\bibitem[{Frost et~al.(2009)}]{bhrot}
\bibinfo{author}{J.~Frost}, et~al.,
\newblock \bibinfo{journal}{JHEP} \bibinfo{volume}{10} (\bibinfo{year}{2009})
  \bibinfo{pages}{014}.
\bibitem[{Dai et~al.(2009)}]{Dai:2009by}
\bibinfo{author}{D.~C. Dai}, et~al.,
\newblock \bibinfo{journal}{arXiv:} \bibinfo{volume}{0902.3577 [hep-ph]}
  (\bibinfo{year}{2009}).
\bibitem[{{The CMS Collaboration}(2011)}]{Khachatryan:2010wx}
\bibinfo{author}{{The CMS Collaboration}},
\newblock \bibinfo{journal}{Phys. Lett.} \bibinfo{volume}{B697}
  (\bibinfo{year}{2011}) \bibinfo{pages}{434--453}.
\bibitem[{{The ATLAS Collaboration}(2008)}]{atlas}
\bibinfo{author}{{The ATLAS Collaboration}},
\newblock \bibinfo{journal}{JINST} \bibinfo{volume}{3} (\bibinfo{year}{2008})
  \bibinfo{pages}{S08003}.
\bibitem[{{The ATLAS Collaboration}(2011{\natexlab{a}})}]{Aad:2011dr}
\bibinfo{author}{{The ATLAS Collaboration}},
\newblock \bibinfo{journal}{Eur.Phys.J.} \bibinfo{volume}{C71}
  (\bibinfo{year}{2011}{\natexlab{a}}) \bibinfo{pages}{1630}.
\bibitem[{{The ATLAS Collaboration}(2011{\natexlab{b}})}]{lumi}
\bibinfo{author}{{The ATLAS Collaboration}},
\newblock \bibinfo{journal}{{}} \bibinfo{volume}{{ATLAS-CONF-2011-116}}
  (\bibinfo{year}{2011}{\natexlab{b}}).
  \bibinfo{note}{{http://cdsweb.cern.ch/record/1376384}}.
\bibitem[{{The ATLAS Collaboration}(2010)}]{bib:atlas-sim}
\bibinfo{author}{{The ATLAS Collaboration}},
\newblock \bibinfo{journal}{Eur. Phys. J C} \bibinfo{volume}{70}
  (\bibinfo{year}{2010}) \bibinfo{pages}{823--874}.
\bibitem[{{GEANT4 Collaboration (S. Agostinelli et al.)}(2003)}]{bib:g4}
\bibinfo{author}{{GEANT4 Collaboration (S. Agostinelli et al.)}},
\newblock \bibinfo{journal}{Nucl. Instr. Meth.} \bibinfo{volume}{A506}
  (\bibinfo{year}{2003}) \bibinfo{pages}{250--303}.
\bibitem[{Frixione and Webber(2002)}]{mcnlo1}
\bibinfo{author}{S.~Frixione}, \bibinfo{author}{B.~Webber},
\newblock \bibinfo{journal}{JHEP} \bibinfo{volume}{06} (\bibinfo{year}{2002})
  \bibinfo{pages}{029}.
\bibitem[{Frixione et~al.(2003)Frixione, Nason, and Webber}]{mcnlo2}
\bibinfo{author}{S.~Frixione}, \bibinfo{author}{P.~Nason},
  \bibinfo{author}{B.~Webber},
\newblock \bibinfo{journal}{JHEP} \bibinfo{volume}{08} (\bibinfo{year}{2003})
  \bibinfo{pages}{007}.
\bibitem[{Nadolsky(2008)}]{cteq}
\bibinfo{author}{P.~M. Nadolsky},
\newblock \bibinfo{journal}{Phys. Rev.} \bibinfo{volume}{D78}
  (\bibinfo{year}{2008}) \bibinfo{pages}{031004}.
\bibitem[{Corcella et~al.(2001)}]{herwig}
\bibinfo{author}{G.~Corcella}, et~al.,
\newblock \bibinfo{journal}{JHEP} \bibinfo{volume}{01} (\bibinfo{year}{2001})
  \bibinfo{pages}{010}.
\bibitem[{Butterworth et~al.(1996)}]{jimmy}
\bibinfo{author}{J.~M. Butterworth}, et~al.,
\newblock \bibinfo{journal}{Z. Phys. C} \bibinfo{volume}{72}
  (\bibinfo{year}{1996}) \bibinfo{pages}{637}.
\bibitem[{Sj{\"{o}}strand et~al.(2006)Sj{\"{o}}strand, Mrenna, and
  Skands}]{pythia}
\bibinfo{author}{T.~Sj{\"{o}}strand}, \bibinfo{author}{S.~Mrenna},
  \bibinfo{author}{P.~Z. Skands},
\newblock \bibinfo{journal}{JHEP} \bibinfo{volume}{05} (\bibinfo{year}{2006})
  \bibinfo{pages}{026}.
\bibitem[{{The ATLAS Collaboration}(2011)}]{Aad:2010ir}
\bibinfo{author}{{The ATLAS Collaboration}},
\newblock \bibinfo{journal}{New J. Phys.} \bibinfo{volume}{13}
  (\bibinfo{year}{2011}) \bibinfo{pages}{053033}.
\bibitem[{Kersevan and Richter-Was(2004)}]{acermc}
\bibinfo{author}{B.~P. Kersevan}, \bibinfo{author}{E.~Richter-Was},
\newblock \bibinfo{journal}{arXiv:} \bibinfo{volume}{0405247 [hep-ph]}
  (\bibinfo{year}{2004}).
\bibitem[{Sherstnev and Thorne(2007)}]{mrst}
\bibinfo{author}{A.~Sherstnev}, \bibinfo{author}{R.~S. Thorne},
\newblock \bibinfo{journal}{arXiv:} \bibinfo{volume}{0711.2473 [hep-ph]}
  (\bibinfo{year}{2007}).
\bibitem[{Mangano et~al.(2003)Mangano, Moretti, Piccinini, Pittau, and
  Polosa}]{alpgen}
\bibinfo{author}{M.~Mangano}, \bibinfo{author}{M.~Moretti},
  \bibinfo{author}{F.~Piccinini}, \bibinfo{author}{R.~Pittau},
  \bibinfo{author}{A.~Polosa},
\newblock \bibinfo{journal}{JHEP} \bibinfo{volume}{307} (\bibinfo{year}{2003})
  \bibinfo{pages}{001}.
\bibitem[{{The ATLAS Collaboration}(2011{\natexlab{a}})}]{muid1}
\bibinfo{author}{{The ATLAS Collaboration}},
\newblock \bibinfo{journal}{{}} \bibinfo{volume}{{ATLAS-CONF-2011-021}}
  (\bibinfo{year}{2011}{\natexlab{a}}).
  \bibinfo{note}{{http://cdsweb.cern.ch/record/1336750}}.
\bibitem[{{The ATLAS Collaboration}(2011{\natexlab{b}})}]{muid2}
\bibinfo{author}{{The ATLAS Collaboration}},
\newblock \bibinfo{journal}{{}} \bibinfo{volume}{{ATLAS-CONF-2011-008}}
  (\bibinfo{year}{2011}{\natexlab{b}}).
  \bibinfo{note}{{http://cdsweb.cern.ch/record/1330715}}.
\bibitem[{{The ATLAS Collaboration}(2011{\natexlab{c}})}]{minbias}
\bibinfo{author}{{The ATLAS Collaboration}},
\newblock \bibinfo{journal}{New J. Phys.} \bibinfo{volume}{13}
  (\bibinfo{year}{2011}{\natexlab{c}}) \bibinfo{pages}{053033}.
\bibitem[{{The ATLAS Collaboration}(2011{\natexlab{d}})}]{Aad:2011sc}
\bibinfo{author}{{The ATLAS Collaboration}},
\newblock \bibinfo{journal}{arXiv:} \bibinfo{volume}{1109.5816}
  (\bibinfo{year}{2011}{\natexlab{d}}). \bibinfo{note}{Submitted to European
  Physical Journal C}.
\bibitem[{Moch and Uwer(2008)}]{Moch:2008qy}
\bibinfo{author}{S.~Moch}, \bibinfo{author}{P.~Uwer},
\newblock \bibinfo{journal}{Phys. Rev.} \bibinfo{volume}{D78}
  (\bibinfo{year}{2008}) \bibinfo{pages}{034003}.
\bibitem[{Langenfeld et~al.(2009)Langenfeld, Moch, and
  Uwer}]{Langenfeld:2009tc}
\bibinfo{author}{U.~Langenfeld}, \bibinfo{author}{S.~Moch},
  \bibinfo{author}{P.~Uwer}  (\bibinfo{year}{2009}).
\bibitem[{Aliev et~al.(2010)}]{topxsec}
\bibinfo{author}{M.~Aliev}, et~al.,
\newblock \bibinfo{journal}{arXiv:} \bibinfo{volume}{1007.1327 [hep-ph]}
  (\bibinfo{year}{2010}).
\bibitem[{{The ATLAS collaboration}(2011)}]{topxsec1fb}
\bibinfo{author}{{The ATLAS collaboration}},
\newblock \bibinfo{journal}{arXiv:} \bibinfo{volume}{1108.3699}
  (\bibinfo{year}{2011}). \bibinfo{note}{Submitted to Physics Letters B}.
\bibitem[{{The ATLAS Collaboration}(2011)}]{Aad:2010ey}
\bibinfo{author}{{The ATLAS Collaboration}},
\newblock \bibinfo{journal}{Eur. Phys. J.} \bibinfo{volume}{C71}
  (\bibinfo{year}{2011}) \bibinfo{pages}{1577}.
\bibitem[{Nason(2004)}]{powheg}
\bibinfo{author}{P.~Nason},
\newblock \bibinfo{journal}{JHEP} \bibinfo{volume}{11546}
  (\bibinfo{year}{2004}) \bibinfo{pages}{040}.
\bibitem[{{The ATLAS Collaboration}(2011)}]{Aad:2011rr}
\bibinfo{author}{{The ATLAS Collaboration}},
\newblock \bibinfo{journal}{arXiv:} \bibinfo{volume}{1109.0525}
  (\bibinfo{year}{2011}). \bibinfo{note}{Submitted to Physics Letters B}.
\bibitem[{Read(2002)}]{cls}
\bibinfo{author}{A.~L. Read},
\newblock \bibinfo{journal}{J. Phys.} \bibinfo{volume}{G28}
  (\bibinfo{year}{2002}) \bibinfo{pages}{2693--2704}.
\bibitem[{Meade and Randall(2008)}]{Meade:2007sz}
\bibinfo{author}{P.~Meade}, \bibinfo{author}{L.~Randall},
\newblock \bibinfo{journal}{JHEP} \bibinfo{volume}{0805} (\bibinfo{year}{2008})
  \bibinfo{pages}{003}.

\end{thebibliography}

\clearpage
\begin{flushleft}
{\Large The ATLAS Collaboration}

\bigskip

G.~Aad$^{\rm 48}$,
B.~Abbott$^{\rm 111}$,
J.~Abdallah$^{\rm 11}$,
A.A.~Abdelalim$^{\rm 49}$,
A.~Abdesselam$^{\rm 118}$,
O.~Abdinov$^{\rm 10}$,
B.~Abi$^{\rm 112}$,
M.~Abolins$^{\rm 88}$,
H.~Abramowicz$^{\rm 153}$,
H.~Abreu$^{\rm 115}$
% ...
\bigskip

% ...
$^{10}$ Institute of Physics, Azerbaijan Academy of Sciences, Baku, Azerbaijan\\
$^{11}$ Institut de F\'isica d'Altes Energies and Departament de F\'isica de la Universitat Aut\`onoma  de Barcelona and ICREA, Barcelona, Spain\\
% ...
$^{48}$ Fakult\"{a}t f\"{u}r Mathematik und Physik, Albert-Ludwigs-Universit\"{a}t, Freiburg i.Br., Germany\\
$^{49}$ Section de Physique, Universit\'e de Gen\`eve, Geneva, Switzerland\\
% ...
$^{88}$ Department of Physics and Astronomy, Michigan State University, East Lansing MI, United States of America\\
% ...
$^{111}$ Homer L. Dodge Department of Physics and Astronomy, University of Oklahoma, Norman OK, United States of America\\
$^{112}$ Department of Physics, Oklahoma State University, Stillwater OK, United States of America\\
% ...
$^{115}$ LAL, Univ. Paris-Sud and CNRS/IN2P3, Orsay, France\\
% ...
$^{118}$ Department of Physics, Oxford University, Oxford, United Kingdom\\
% ...
$^{153}$ Raymond and Beverly Sackler School of Physics and Astronomy, Tel Aviv University, Tel Aviv, Israel\\
% ...
\end{flushleft}


\end{document}